\documentclass{article}
\usepackage{color}
\usepackage{graphicx, graphics}
\usepackage{subfigure}
\usepackage{amsmath,amstext}  % pour utiliser amslatex
\usepackage{amsfonts,amssymb, bbm} % polices et symboles AMS
\usepackage[LGR, T1]{fontenc}
\usepackage[utf8]{inputenc}
\usepackage[english]{babel}
\usepackage{epsfig}
\usepackage{epsf}
\usepackage{vmargin}
\newcommand{\cY}{\mathcal{Y}}
\newcommand{\cG}{\mathcal{G}}
\newcommand{\R}{\mathbb{R}}

\def\1{1\kern-.20em {\rm l}}

\title{Hidden Markov Model for the detection of a degraded operating mode of optronic equipment} 
\author{ Camille Baysse, Anne G\'egout-Petit, Jérôme Saracco\\
Universit\'e de Bordeaux, IMB CNRS UMR~5251\\
and INRIA~Bordeaux~Sud~Ouest team CQFD, France\\
\and Didier Bihannic, Michel Prenat \\
Thales Optronique}

\begin{document}
\maketitle
  
  \begin{abstract}
As part of optimizing the reliability, Thales Optronics now includes systems that examine the state of its equipment. The aim of this paper is to use hidden Markov Model to detect as soon as possible a change of state of optronic equipment in order to propose maintenance before failure. For this, we carefully observe the dynamic of a variable called "cool down time" and noted Tmf, which reflects the state of the cooling system. Indeed, the Tmf is an indirect observation of the hidden state of the system. This one is modelled by a Markov chain and the Tmf is a noisy function of it. Thanks to filtering equations, we obtain results on the probability that an appliance is in degraded state at time $t$, knowing the history of the Tmf until this moment. We have evaluated the numerical behavior of our approach on simulated data. Then we have applied this methodology on our real data and we have checked that the results are consistent with the reality. This method can be implemented in a HUMS (Health and Usage Monitoring System). This simple example of HUMS would allow the Thales Optronics Company to improve its maintenance system. This company will be able to recall appliances which are estimated to be in degraded state and do not control to soon those estimated in stable state.
\end{abstract}

%==============
\section{Introduction}
%==============

Thales Optronics aims to optimize the ratio availability - cost. The company wants to reduce the failure rate of these appliances by the evolution of its maintenance concept which passes from a logic of repair to a logic of anticipation of these defects. As part of optimizing the reliability, Thales Optronics now includes systems that examine the state of its equipment. This function is performed by HUMS (Health and Usage Monitoring System).
The role of HUMS is :
\begin{enumerate}
\item to record environmental conditions and use of equipment,
\item to evaluate the state of the system,
\item to anticipate and alert about the excesses of operation,
\item to optimize maintenance operations.
\end{enumerate}
Our approach comes within a specific context. In this paper, we focus on point 2. We have at our disposal a variable that reflects the state of the system and we want to detect a change in mode of this variable (which is a change of slope in our case). There exist different methods for this kind of detection as the CUSUM, presented for instance by Basseville \& al in \cite{M.Basseville}. But in this paper we focus on hidden Markov chains to detect this change of mode. The state of our system at time $t$ is then modeled by a Markov chain $X_{t}$. In our case we do not observe directly this chain but indirectly through the Tmf variable, a noisy function of this chain. We will see in this paper how we can address this issue by using filtering theory.

For this, we will first introduce the industrial problem in section $2$ and the mathematical model in a general case in section $3$. Section $4$ presents a simulation study and section $5$ the implementation of the methodology on our real data. 
\section{Industrial problem}
Each of the appliances has a logbook which provides the following information at each start-up: number of uses, cumulative operating time of appliance, initial temperature and the ``cool down time'' (Tmf for ``temps de mise en froid'' in french). This Tmf is the transit time for the system from ambient temperature to a very low one. This temperature decrease is required to operate appliance and this is done on every boot. According to experts, a Tmf increase results from deterioration in the system. According to this hypothesis, a careful observation of Tmf evolution would allow us to determine the state of the system and prevent the breakdown. So we will look at the evolution of Tmf which seems to be a good indicator of the system state.

We suppose that the system has three possible states:
\begin{itemize}
\item Stable state: Tmf is constant. This reflects a system in good working order. There is no anomaly to report.
\item Degraded state: the Tmf increases. This reflects a specific deterioration in the system.
\item Failure: the system is stopped.
\end{itemize}

Appliances move from stable state to degraded state, to reach failure. It is important to detect the beginning of a degradation to prevent as soon as possible occurrence of failure.
As explained above, we would not observe directly the state of our system but indirectly through the Tmf. Our objectives are:
\begin{itemize}
\item to estimate at every moment the state of the system by the evaluation of the probability of being in degraded state knowing the history of Tmf until this moment,
\item to detect as soon as possible the degradation of the system for a maintenance action before failure.
\end{itemize}

To solve this problem, we use hidden Markov chains.

%=========================
\section{Hidden Markov model}
%=========================

In this section, we provide a general mathematical framework to tackle our problem. We have to detect a rupture in the behavior of the variable Tmf. There exist different methods for this kind of detection (see for instance Basseville and Nikiforov \cite{M.Basseville}). We choose to use Hidden Markov Model (HMM). HMM are frequently used to detect point mutations in DNA in genomics (see for instance Fridlyand \cite{J.Fridlyand}) or in speech recognition (see Rabiner \cite{L.R.Rabiner}). In the domain of reliability, HMM are also used in a context of high frequencies data (see Wang \cite{W.Wang}). In our context, the size of our data is not large (28 appliances, maximum 400 recordings in a logbook). But in the following, we will see that this tool is also powerful in our context. We first present the model in a general case and the estimation of the parameters of interest.

%-------------------------------
\subsection{Modeling}
%-------------------------------

\subsubsection{Main process}
%------------------------------------------

Consider $(X_{t})_{t>0}$ a Markov chain in continuous time, defined on a probability space ($\Omega$, $F$, $P$) with discrete state space S=$\left\{e_{1},e_{2},\dots,e_{N}\right\}\in\R^{N}$. So $X_{t}=(X_{t}^{1},\dots,X_{t}^{N})$ is a vector of $\R^{N}$. For convenience, we follow Elliott's assumptions \cite{Elliott} and we set $e_{i}= (0, 0,\dots, 1, 0,\dots, 0)$ so that $(e_{1},e_{2},\dots,e_{N})$ is an orthonormal basis of $\R^{N}$.

Let us denote the probability $p_{t}^{i}=P(X_{t}=e_{i})$ for $0\leq i \leq N$ and $p_{t}=(p_{t}^{1},\dots,p_{t}^{N})$. The motion of the chain $X_{t}$ depends on $A=(a_{ij})$, the Q-matrix of the process (see C.Cocozza-Thivent \cite{Cocozza} for definition). The vector $ p_{t}$ is linked to matrix $A$ by the Kolmogorov equation $\frac{dp_{t}}{dt}=A$ and $X_{t}$ has the semimartingale representation:
\begin{equation}
X_{t} = X_{0} + \int_{0}^{t}AX_{r}dr + V_{t}
\end{equation}
with $V_{t}$ a martingale.

\subsubsection{Observation process}
%--------------------------------------------------

$X_{t}$ is not directly observed, but through the process $Y_{t}$ given by the formula:
\begin{equation}\label{eq.simu}
Y_{t} = \int_{0}^{t}c(X_{r})dr + W_{t}
\end{equation}
with:
\begin{itemize}
\item $(W_{t})_{t>0}$ a standard Brownian motion on $(\Omega,F,P)$ independent of $(X_{t})_{t>0}$,
\item $c(X_{t})=<X_{t};c>$ where $<;>$ is the scalar product in $\R^{N}$ and $c=(c_{1},\dots,c_{N})\in \R^{N}$.
\end{itemize}
So, in mean, the increase of the observed process $Y_{t}$ depends on the state of $X_{t}$ through $c(X_{t})$. 
A Brownian noise is added to the slope $c(X_{t})$. 

Let us denote:
\begin{itemize}
\item $(\cY_{t})_{t\geq0}$ the right-continuous complete filtration generated by $\sigma(Y_{s}: 0\leq s\leq t)$,
\item  $(\cG_{t})_{t>0}$ the right-continuous complete filtration generated by $\sigma (X_{s},Y_{s}:0 \leq s \leq t)$.
\end{itemize}

Recall that our aim is to determine the probability of the system to be in a particular state knowing the trajectories of $Y$ until $t$. The best $L^{2}$-approximation of this quantity is given by the conditional probability
$\hat{p}^{i}_{t}=\textsl{P}(X_{t}= e_{i}\left|\cY_{t}\right.)$ for $0\leq i\leq N$.
Note that 
$$\textsl{P}(X_{t}= e_{i}\left|\cY_{t}\right. )=\textsl{P}(X_{t}^{i}=1\left|\cY_{t}\right.)=E\left[ X_{t}^{i}|\cY_{t}\right]=\left(E\left[ X_{t}|\cY_{t}\right]\right)_{i}.$$
So we have to compute the $N$-dimensional conditional expectation $E\left[ X_{t}|\cY_{t}\right]$. This is the aim of the next section.

%-----------------------------------------------------------------------------
\subsection{Filtering equations and parameters estimation}
%------------------------------------------------------------------------------

First, we give filtering equations which provide conditional expectations of functions of $X_{t}$, knowing the story of $Y_{t}$. Then we will see how these equations allow us to estimate parameters $A$, $c$ and the probability of being in a state $e_{i}$ given $\cY_{t}$.

\subsubsection{Filtering equations}
%-------------------------------------------------

Elliott \cite{Elliott} gives unnormalized filtering equations of different fonctionnals of $X_{t}$. To write these equations, let us denote $\sigma(F(X_{s}, s\leq t))=\bar{\textsl{E}}\left[\bar{\wedge}_{t}F(X_{s}, s\leq t)\left| \cY_{t} \right. \right]$ with $\bar{P}$ and $\bar{\wedge}_{t}$ associated with the absolutely continuous probability of change:
$$\left.\frac{dP}{d\bar{P}} \right|_{\cG_t} = \bar{\wedge}_{t}
= \exp \left(\int_{0}^{t} < c ; X_{r} > dY_{r} - \frac{1}{2} \int_{0}^{t} < c ; X_{r} >^{2} dr\right).$$
This change of probability is a standard method in filtering because under $\bar{P}$, $Y_{t}$ is independent of $X_{t}$. Under $\bar{P}$, the dynamic of unnormalized filter satisfies stochastic differential equation.

Filtering equations are about:
\begin{itemize}
\item state of the system:
$$\sigma(X_{t})=\sigma(X_{0}) + \int^{t}_{0}A\sigma(X_{r})dr+\int^{t}_{0}C\sigma(X_{r})dY_{r},$$
\item number of jumps from $e_{i}$ to $e_{j}$ in the time interval $\left[0,t\right]$ denoted $\varsigma^{ij}_{t}$:
$$\sigma(\varsigma^{ij}_{t}X_{t})=\int^{t}_{0}<\sigma(X_{r});e_{i}>a_{ji}e_{i}dr+\int^{t}_{0}A\sigma(\varsigma^{ij}_{r}X_{r})dr+\int^{t}_{0}C\sigma(\varsigma^{ij}_{r}X_{r})dY_{r},$$
\item waiting time in state $e_{i}$ on the interval $\left[0,t\right]$ denoted $\vartheta^{i}_{t}$:
$$\sigma(\vartheta^{i}_{t}X_{t})=\int^{t}_{0}<\sigma(X_{r});e_{i}>e_{i}dr+\int^{t}_{0}A\sigma(\vartheta^{i}_{r}X_{r})dr+\int^{t}_{0}C\sigma(\vartheta^{i}_{r}X_{r})dY_{r},$$
\item drift defined by $T^{i}_{t}=\int^{t}_{0}<X_{r};e_{i}>dY_{r}$:
$$\sigma(T^{i}_{t}X_{t})=c_{i}\int^{t}_{0}{\sigma(<X_{r},e_{i}>e_{i})dr}+\int^{t}_{0}{A\sigma(T^{i}_{t}X_{r})dr}+\int^{t}_{0}\left[{<\sigma(X_{r});e_{i}>e_{i}+C\sigma(T^{i}_{r}X_{r})}\right]dY_{r}. $$
\end{itemize}

These equations about $\varsigma^{ij}$, $\vartheta^{i}$, $T^{i}$ are useful for the estimation of $A$ and $c$ when $Y_{t}$ is observed in a long time.

\subsubsection{Estimation}
%----------------------------------------------------------------------------------------------------------------------------------------------

Maximum likelihood estimation of $A$ and $c$ leads to the estimators:
\begin{equation}\label{eq.A.c}
\hat{a}_{ij}(t)=\frac{\sigma(\varsigma^{ij}_{t})}{\sigma(\vartheta^{i}_{t})} \mbox{~~and~~} \hat{c}_{i}(t)=\frac{\sigma(T^{i}_{t})}{\sigma(\vartheta^{i}_{t})}.
\end{equation}
These estimators converge with $t$ according to R.J. Elliott \cite{Elliott}. Using filtering equations and the estimators of $A$ and $c$, it is possible to compute the estimated probability of the system to be in state $e_{i}$ thanks to the following formula:
\begin{equation}\label{eq.proba}
\textit{E} \left[X_{t} \left|\cY_{t} \right. \right]=\frac{\sigma(X_{t})}{\sigma(1)}.
\end{equation}
Indeed, $\textsl{P}(X_{t}= e_{i}|\cY_{t})=(E [X_{t}|\cY_{t}])_{i}=(\frac{\sigma(X_{t})}{\sigma(1)})_{i}$.

Note that filtering equations don't give directly $\sigma(\varsigma^{ij}_{t})$, $\sigma(T^{i}_{t})$, $\sigma(\vartheta^{i}_{t})$ and $\sigma(1)$ but $\sigma(\varsigma^{ij}_{t}X_{t})$, $\sigma(T^{i}_{t}X_{t})$, $\sigma(\vartheta^{i}_{t}X_{t})$ and $\sigma(X_{t})$. To pass from one to another, we just have to multiply these elements by vector $(1,1,..,1)^{T}$ to obtain $\sigma(\varsigma^{ij}_{t})$, $\sigma(T^{i}_{t})$, $\sigma(\vartheta^{i}_{t})$ and $\sigma(1)$. Indeed, thanks to assumptions on $X_{t}\in(e_{1},\dots,e_{N})$, $\left\langle X_{t} ; (1,1,..,1) \right\rangle =1$.

Let us now illustrate this approach on simulated data.

%=====================
\section{Simulation study}
%=====================

In this section, the framework is the following. We assume that the process has two possible states: first $X_{t}=e_{1}$ and second $X_{t}=e_{2}$ with transitions from $e_{1}$ to $e_{2}$ and conversely. For instance, $e_{1}$ (respectively $e_{2}$) could correspond to the stable state (respectively the degraded state). So $X_{t}$ oscillates between these two states.

%-----------------------------------------------------------------------
\subsection{Probability estimation of being in a degraded state}
%-----------------------------------------------------------------------

We first suppose that we know matrix $A$ and vector $c$. The component $a_{12}$ of $A$ is the parameter of the exponential distribution of the time in stable state (before degraded state) and $a_{21}$ is the parameter of the exponential distribution of the time in degraded state. Using these values, we can simulate $X_{t}$. Then, using values of $c$ and $X_{t}$, we simulate $Y_{t}$ thanks to equation (\ref{eq.simu}) and Euler scheme approximation to simulate stochastic differential equations. Now that our data are simulated and our parameters known we can compute the conditional probability that the system is in degraded state. For this, we use equation (\ref{eq.proba}) for the computation of the conditional probabilities $\hat{p}^{2}_{t}$ and $\hat{p}^{1}_{t}=1-\hat{p}^{2}_{t}$. This computation is made again by a recursive algorithm that uses the Euler scheme to approximate stochastic differential equations.

\begin{figure}[!htb]
\begin{center}
\includegraphics[width=0.85\textwidth]{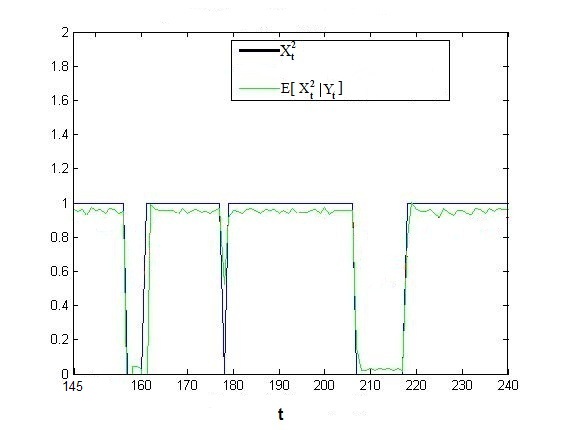}	
\caption{\label{fig1}Evolution of $X_{t}^{2}$ and estimation of conditional probability $P\left(X_{t}^{2}=1\left|\cY_{t}\right.\right)$}
\end{center}
\end{figure}

An illustration of the good numerical behavior of the computational process is given in Figure~\ref{fig1}. This figure zooms on a part of the trajectory of $X^{2}_{t}$ and $\hat{p}^{2}_{t}=E\left[X^{2}_{t}\left|\cY_{t}\right.\right]$. We clearly observe that the filter correctly provides the evaluation of $X^{2}_{t}$ that is close to 1 (respectively 0) when $X^{2}_{t}=1$ (respectively when $X^{2}_{t}=0$).

Note that in this simulation study, we assume that parameters $A$ and $c$ are known. This is not the case in practice and these parameters must be estimated before estimating probability $\hat{p}^{1}_{t}$ and $\hat{p}^{2}_{t}$.

%--------------------------------------------------
\subsection{Parameters estimation}
%--------------------------------------------------

\subsubsection{Estimation of matrix $A$ and vector $c$}
%----------------------------------------------------------------------------
With simulations of process $X_{t}$ in a long time, it is possible to use formula (\ref{eq.A.c}) to estimate parameters $A$ and $c$.

We first suppose vector $c$ known and we seek to estimate the matrix $A$ from observations of $Y_{t}$. However, one difficulty of this estimation step is the fact that $\sigma(\varsigma^{ij}_{t})$ and $\sigma(\theta^{i}_{t})$ are governed by $A$. So we developed an iterative algorithm to approximate $A$ starting with an arbitrary $A_{0}$, operating in the following way: at step $k$, we use $\hat{A}_{k-1}$ to compute $\hat{A}_{k}$ via filtering equation (\ref{eq.A.c}). The convergence of this estimator has been proved by Zeitouni and Dembo \cite{dembo}.

Now we assume matrix $A$ known and we seek to estimate vector $c$ from observations of $Y_{t}$. For this, we also use formula (\ref{eq.A.c}). Once again, $\sigma(T^{i}_{t})$ and $\sigma(\theta^{i}_{t})$ are governed by $c$ itself. So, by the same method as previously, we developed again an iterative algorithm to approximate $c$ starting with an arbitrary vector $c_{0}$.

%----------------------------------------------------------------------------------------------------------------------------------------------------
\subsubsection{Sensitivity of filter $P(X_{t}=e_{2}\left|\cY_{t}\right.)$ to parameters $A$ and $c$}\label{sect.sensitivity}
%----------------------------------------------------------------------------------------------------------------------------------------------------

Since the values of $A$ and $c$ are unknown in practice, it seems important to study the impact of a poor estimation of $A$ and $c$ in calculation of probability $P (X_{t}=e_{2}\left|\cY_{t}\right.)$.
We simulate $X_{t}$ and $Y_{t}$ for given values of $A$ and $c$:  $A=
\begin{pmatrix}
-0.1&0.1\\
0.05&-0.05\\
\end{pmatrix}$
and $c= (-1; 1)$. We then estimate probability of being in a degraded state.

\begin{figure}[!htb]
\begin{center}
\includegraphics[width=0.8\textwidth]{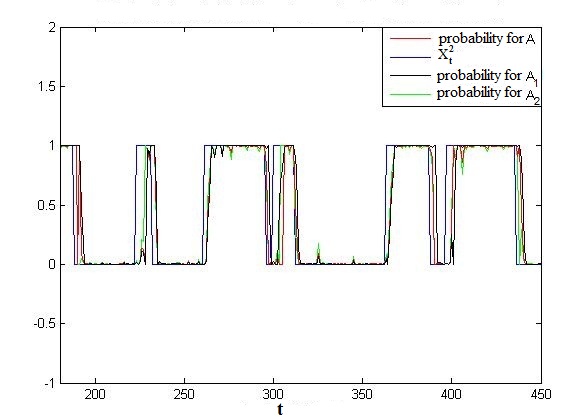}
\caption{\label{fig2} Evolution of probability of being in a degraded state for different matrices A}
\end{center}
\end{figure}

In a first step, we consider small deviations from the true matrix $A$: $A_{1}=
\begin{pmatrix}
-0.01&0.01\\
0.04&-0.04\\
\end{pmatrix}$ 
and $A_{2}=
\begin{pmatrix}
-0.2&0.2\\
0.08&-0.08\\
\end{pmatrix}$. With these two matrices, we again compute probability of being in a degraded state. Figure~\ref{fig2} gives these estimations. We clearly observe that deviations do not severely impact on the probability of interest.

\begin{figure}[!htb]
\begin{center}
\includegraphics[width=0.8\textwidth]{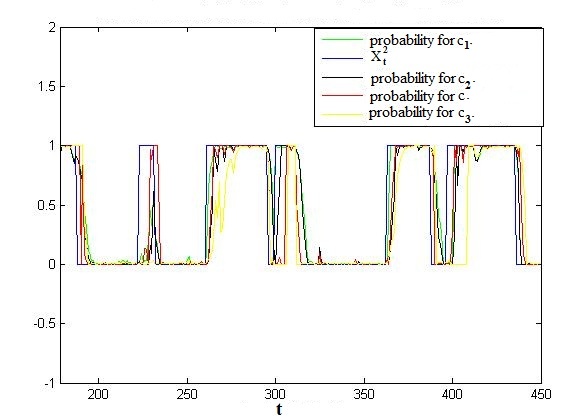}
\caption{\label{fig3} Evolution of probability of being in a degraded state for different vectors c}
\end{center}
\end{figure}

In a second step, we estimate this probability with deviations from the true value $c$: $c_{1}= (-0.5, 0.5)$, $c_{2}= (-1, 0.5)$ et $c_{3}= (0, 1)$. Figure~\ref{fig3} gives these corresponding estimations. Again we observe that deviations do not severely impact the probability of interest.

In our real data study we do not estimate parameters $A$ and $c$  by the method given in section $4.2.1$, indeed the process is stopped at its first transition to a degraded state and we do not observe $X_{t}$ in long time with many transitions as in the simulation study. But from the previous simulations, we have noticed that a misspecification of these parameters does not seem to strongly impact on the filter value $P(X_{t}=e_{2}\left|\cY_{t}\right.)$.

%=============================
\section{Application to industrial case}
%=============================

%-----------------------
\subsection{Data}
%-----------------------

We have logbooks of 28 appliances: five of them failed at the end of the study, due to a mechanical malfunction in the cooling system. For other appliances, the failures were not mechanical and are considered to be unpredictable (not related to a degradation effect and often due to an electronic failure). From the logbooks, we recover Tmf value and initial temperature at each startup of the system. The time unit of the model is the number of startups and we assume a common model for all appliances ($A$ and $c$ are the same for all of them) and the motion of the 28 appliances are mutually independent.

%-----------------------
\subsection{Preliminary data processing}
%-----------------------

\begin{figure}[!htb]
\begin{center}
\includegraphics[width=1.00\textwidth]{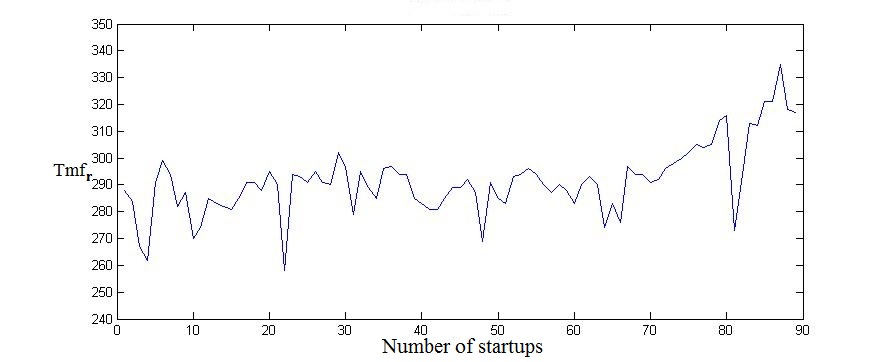}\\
\includegraphics[width=1.00\textwidth]{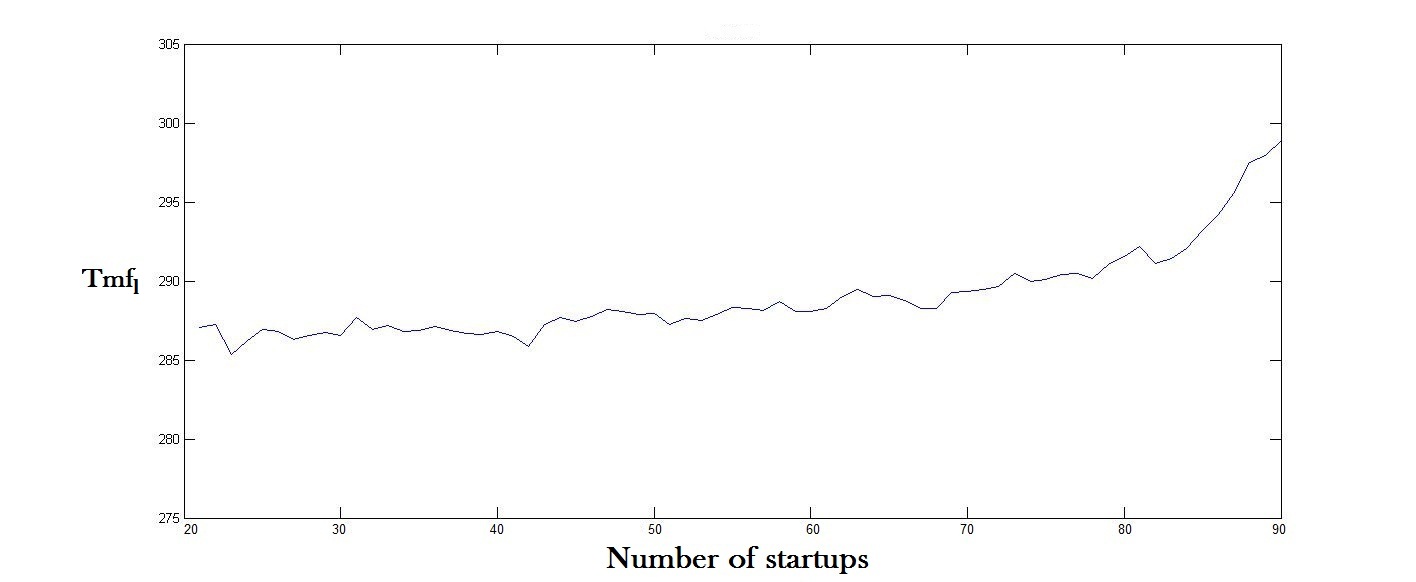}
\caption{\label{fig4}Evolution of corrected Tmf over uses at $10^o$ C and smoothing data associated}
\end{center}
\end{figure}

The two variables, Tmf and initial temperature are linked together. Indeed, a high (resp. low) initial temperature increases (resp. decreases) the Tmf. So it was necessary to correct this crude Tmf by a standard linear regression according to initial temperature of appliance. We use this regression to bring the Tmf to a setting where initial temperature is constant and equals $10^{\circ}$C. This corrected Tmf is denoted by $Tmf_{r}$ in the following. In Figure~\ref{fig4}, we provide the corrected Tmf evolution of one appliance. We can see a very noisy phenomenon. Down peaks may be the result of ``on/off/on'' too brutal for appliance: the system is on, turned off and back on instantly so that initial temperature remains low. To soften this phenomenon, we decide to smooth the corrected Tmf ($Tmf_{r}$). For this, we compute a moving average of $Tmf_{r}$ as follows:
$$Tmf_{l}(j)=\frac{\sum_{i=j}^{20+j-1} Tmf_{r}(i)}{20},$$
where $Tmf_{r}(i)$ is the value of corrected Tmf at the $i^{th}$ startup. Let us denote by $Tmf_{l}$ the smoothing correcting Tmf value. In our modeling, we set $Y_{t}=Tmf_{l}(t)$. A theoretical interest of this smoothing step is that the filtering method works well with a not too noisy signal. Note that the $Tmf_{l}$ starts at the $20^{th}$ startup because it is necessary to have $20$ $Tmf_{r}$ to compute $Tmf_{l}$. In practice it is necessary to wait $20$ startups before the first computation of the probability of being in a degraded state. In Figure~\ref{fig4}, we plot the evolution of smooth corrected Tmf of this appliance.

We can notice on the bottom graph of Figure~\ref{fig4} that $Tmf_{l}$ remains constant for a while and then gradually increases. This change in slope was not obvious in the top graph of Figure~\ref{fig4}. This is another interest of the smoothing step. Now, from these $Tmf_{l}$ values, we are able to calculate probability of being in degraded state. For this, we first need to estimate parameters $A$ and $c$.

%-----------------------------------------------------------------------
\subsection{Estimation of parameters A and c}
%-----------------------------------------------------------------------

The estimation method presented in section 4.2.1 using the observation of the process in a long time is not possible here. Indeed the real system does not oscillate between two states because it is stopped at its first transition to degraded state. Then we propose a practical choice for $c$ and $A$ mixing estimation using the data from the $28$ appliances and expert opinions.

According to experts, slope of smoothed curve is close to 0 when the system is in stable state and it is strictly positive when it degrades. In addition, according to graphs of the evolution of the $Tmf_{l}$, the slope is close to 1 when the system is in a degraded state. So we can naturally set $\hat{c}=(0, 1)$.

About the $Q$-matrix $A=
\begin{pmatrix}
-a_{12}&a_{12}\\
a_{21}&-a_{21}\\
\end{pmatrix}$, $a_{12}$ is the parameter of the exponential distribution of the time in stable state (before degraded state). We have estimated this parameter using our data (28 appliances: $5$ times of failure and $23$ censures). By standard survival method taking censures into account, we have first estimated $a_{12}$ by $\frac{1}{1000}$. In order to detect as soon as possible a change of state (contraint requested by Thales) and according to our study of sensitivity (see Section~\ref{sect.sensitivity}), we chose to put a value $10$ times greater: $\hat{a}_{12}=\frac{1}{100}$.

The coefficient $a_{21}$ should equal zero because system in degraded state can not return to a stable state. But in our equations, our filter $P(X_{t}=e_{2}\left|\cY_{t}\right.)$ must be versatile, so we have chosen a small value $\hat{a}_{21}=\frac{1}{1000}$. With this choice, the chance that an appliance in degraded state comes to stable state is very small.

Now, we are able to estimate probability of interest.

%--------------------------------
\subsection{Results}
%--------------------------------

\begin{figure}[!htb]
\begin{center}
\includegraphics[width=1.00\textwidth]{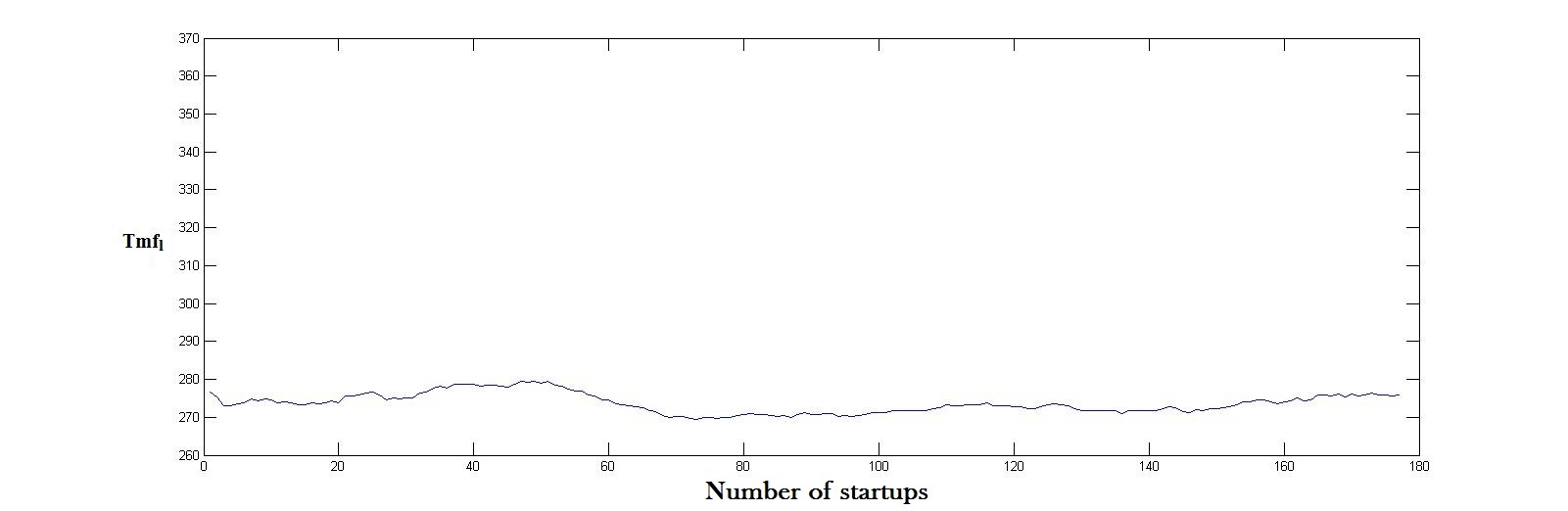}\\
\includegraphics[width=1.00\textwidth]{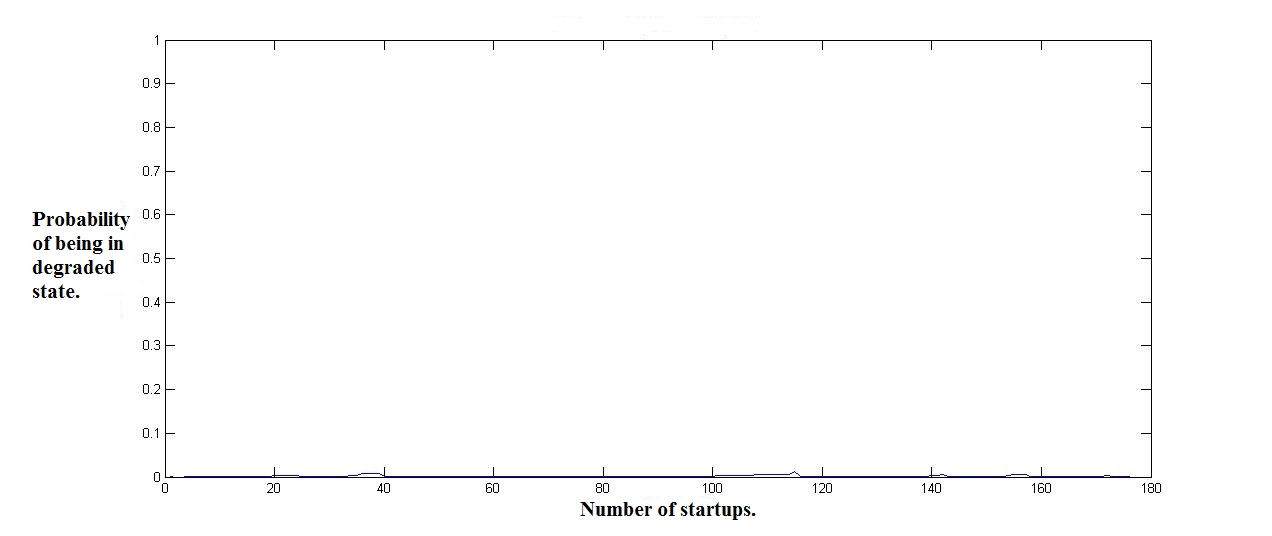}
\caption{\label{fig5} Smoothing data and evolution of probability to be in a degraded state for $E_{h}$}
\end{center}
\end{figure}

With this choice for $A$ and $c$ and using the filtering equation ($4$), we computed the probability of being in a given state, at each startup $t$, knowing the story $\cY_{t}$.
We first consider appliance noted $E_{h}$. A posteriori, we can see that $E_{h}$ was trouble-free during its whole history. Figure~\ref{fig5} gives the evolution of its $Tmf_{l}$. At each time $t$, we compute its probability of being in degraded state through (\ref{eq.proba}) using values of $Tmf_{l}$ before $t$. We clearly observe a $Tmf_{l}$ quite constant during uses and a probability of being in a degraded state close to zero.

\begin{figure}[!htb]
\begin{center}
\includegraphics[width=1.00\textwidth]{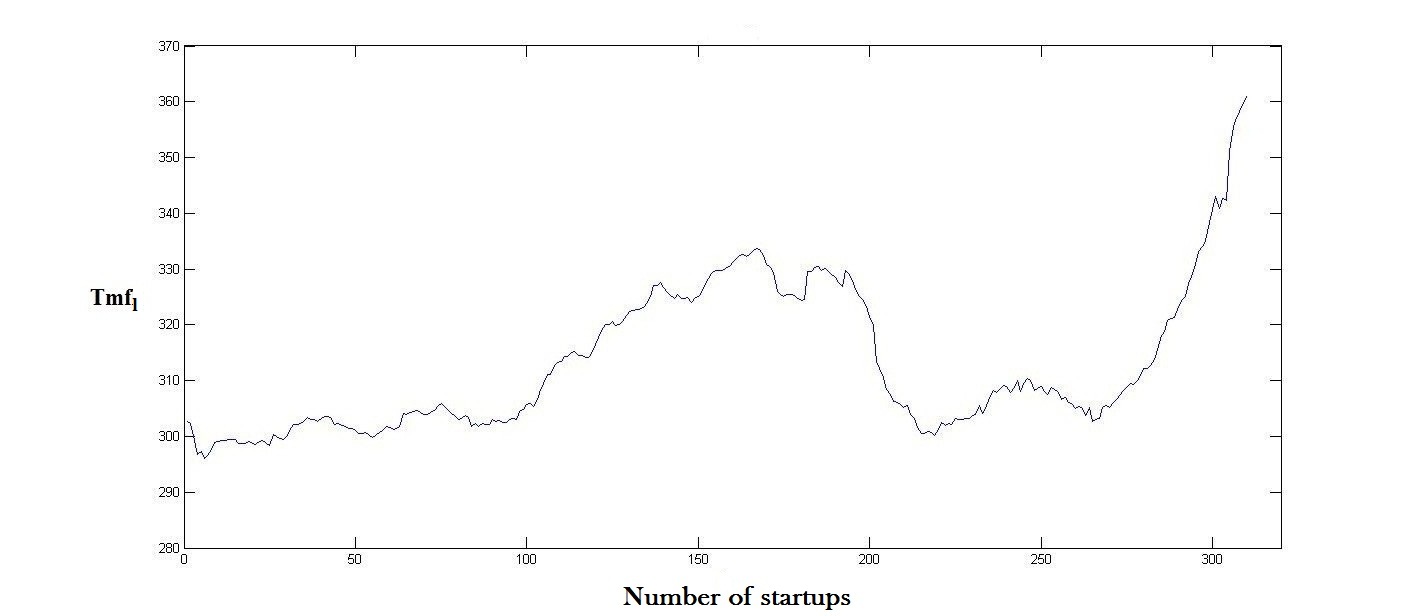}
\includegraphics[width=1.00\textwidth]{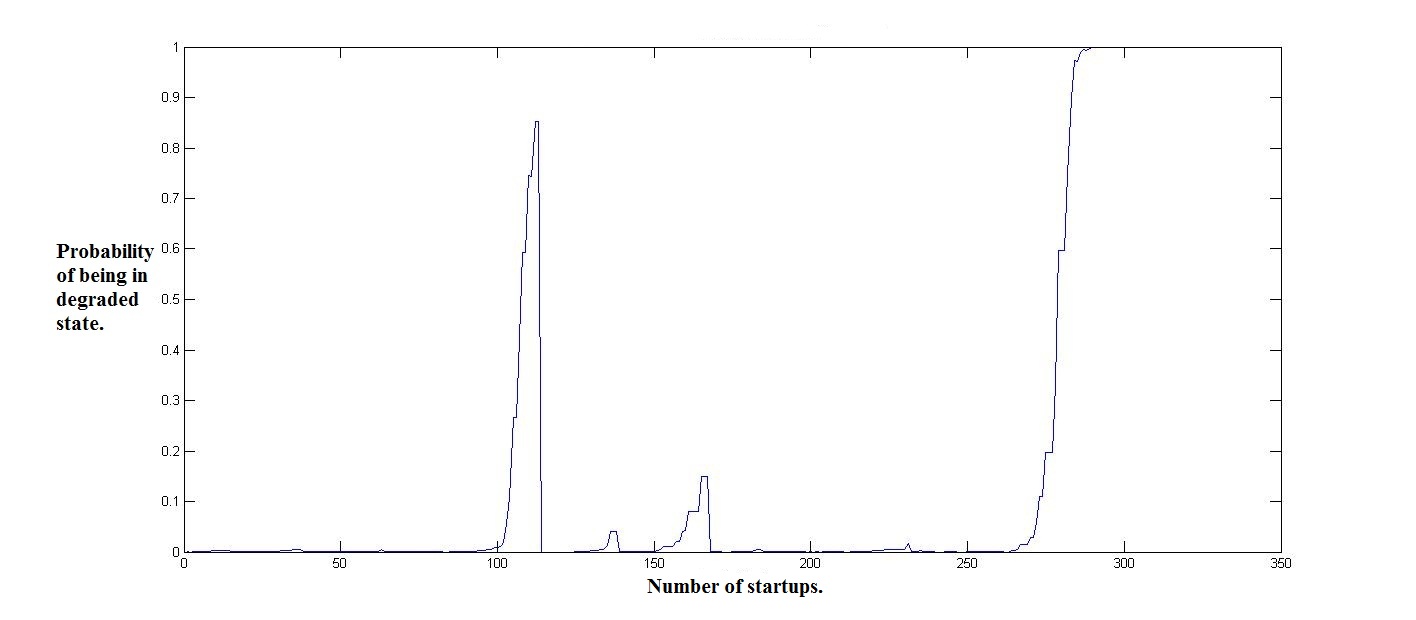}
\caption{\label{fig6} Smoothing data and evolution of probability to be in a degraded state for $E_{d}$}
\end{center}
\end{figure}

Now, we consider an appliance denoted $E_{d}$. A posteriori, we see that $E_{d}$ degrades and breaks down at the end of the study. In Figure~\ref{fig6}, we see a $Tmf_{l}$ quite constant during the first uses; then, $Tmf_{l}$ increases and then decreases to return to starting level. Finally, we notice an abrupt rise of $Tmf_{l}$. Simultaneously, we note that the computed probability of being in degraded state is very low when $Tmf_{l}$ is constant and then sharply increases with $Tmf_{l}$ to one.

To conclude, these two examples illustrate a good numerical behavior of the proposed approach.
Now we are interested in a decision criterion that allows us to detect as soon as possible a degraded state in order to return appliances to perform maintenance action before failure.

%====================
\section{Decision criterion}
%====================

The increase of the probability of being in a degraded state is not sufficient to detect a future failure. We have to propose a decision criterion for maintenance. For this, we have tested different rules based on the fact that probability has to cross a threshold during a number of consecutive uses. We have tried different thresholds combined with different numbers of crossing. At each time, we have recorded the number of false and good detections. It is important to limit both false-positive and false-negative detections. According to the comparison of these rules, we have chosen the following criterion: when the probability to be in degraded state equals 1 over a period of three uses, the appliance is sent back for maintenance.

We applied this rule on our $28$ appliances and we obtain the results presented in Table~\ref{tab1}.

\begin{table}
\begin{tabular}{|l|c|r|}
\hline
Decision criterion & Observed failure & No observed failure\\ 
\hline
Futur failure detected & \textbf{3} & 0 \\
 \hline
Futur failure not detected & 2 & \textbf{23} \\ 
\hline
Total &5&23\\ 
\hline
\end{tabular}
\caption{\label{tab1} Results obtained with the decision criterion}
\end{table}

The decision criterion provides 26 good detections over 28. It does not provide false detection: the 23 appliances without observed failure were not detected as degraded. For the five appliances that failed during the study, the criterion identifies three of them as degraded (before failure). We suppose that for the two appliances which have not been correctly identified, the failure may be not related to a degradation effect of the cooling system and then can not be detected by our proposed approach.

%=====================
\section{Concluding remarks}
%=====================

Using this model, Thales is now working to implement in its HUMS in operating system a new maintenance algorithm.

There are two technical solutions according to the system embedded calculator:
\begin{itemize}
\item system capitalizes data, assesses and provides information about the cooler state,
\item system capitalizes data but the cooler state is assessed by a maintenance laptop plugged periodically on its maintenance socket.
\end{itemize}
This model will allow us to improve the maintenance and the usage policies of monitored system. The improvements are:
\begin{itemize}
\item moving from a preventive or corrective maintenance to a predictive maintenance, this evolution allows to reduce the support cost,
\item ability to create a degraded operational mode,
\item increase the mission success probability (systems will be chosen according to their real status for critical mission).
\end{itemize}

The performance of the new maintenance policy is possible thanks to combination of mathematical, high technology and new maintenance organization.

%=========================

\bibliographystyle{plain}
\bibliography{biblio}

%=========================

\end{document}